\documentclass[multphys,vecphys]{svmult}

\usepackage{makeidx}   
\usepackage{graphicx}  
\usepackage{multicol}  
\usepackage{cite}
\usepackage{epsfig}
\makeindex             

\begin{document}
\title{Quenched Disorder Distributions in Three-Dimensional Diluted 
Ferromagnets} 
\toctitle{Quenched Disorder Distributions in Three-Dimensional Diluted 
Ferromagnets}
%
\titlerunning{Quenched Disorder Distributions in 3D
Diluted Ferromagnets}
%
\author{
W. Janke\inst{1},
P.-E. Berche\inst{2},
C. Chatelain\inst{3}\and
B. Berche\inst{3}
}
%
%
%
\institute{
Institut f\"ur Theoretische Physik, Universit\"at Leipzig,
Augustusplatz 10/11, D-04109 Leipzig, Germany\\
\texttt{wolfhard.janke@itp.uni-leipzig.de}
\and
Groupe de Physique des Mat\'eriaux, Universit\'e de Rouen, 
F-76801 Saint-Etienne du Rouvray Cedex, France\\
\texttt{pierre.berche@univ-rouen.fr}
\and
Laboratoire de Physique des Mat\'eriaux,
Universit\'e Henri Poincar\'e, Nancy I, BP 239,
F-54506 Vand\oe uvre les Nancy Cedex, France\\
\texttt{chatelai, berche@lpm.u-nancy.fr}
}
\maketitle              

\begin{abstract}
We report on large-scale Monte Carlo simulations of the three-dimen\-sional
$4$-state Potts ferromagnet subject to quenched, random
bond dilution. For small dilutions the rather strong first-order phase
transition of the pure system is found to persist, whereas for larger
dilutions the theoretically expected softening to a continuous transition
is confirmed.
The properties of the underlying disorder distributions of thermal
observables are discussed and illustrated with a few selected examples.

\end{abstract}

\section{Introduction}
\label{sec:1}  
Physical systems with quenched, random impurities display under certain
conditions a completely different behaviour than the pure systems. One (often
unwanted) experimental realization are point defects which may be modelled
theoretically by random site dilution \cite{Folk2001}. If the pure system
exhibits a continuous phase transition, the influence of quenched, random
disorder may drive the system into a new universality class, provided the
critical exponent $\alpha_{\rm p}$ of the pure system's specific heat is
positive (Harris criterion) \cite{Harris74}. Also in the case of a 
first-order phase transition in the pure system, quenched disorder can have
a dramatic effect: Phenomenological renormalization group arguments suggest
the possibility of a softening to a continuous transition \cite{ImryWortis79}.
While in two dimensions this effect has been theoretically 
proven \cite{Aizenman89} and numerically confirmed \cite{BeCh02} for any small 
amount of disorder, in three dimensions (3D) one expects 
that with increasing disorder the strength of the transition is gradually 
weakened until beyond
a (tricritical) concentration of impurities a real softening to a 
second-order transition sets in \cite{cardy,cardy_statphys}.

\begin{figure}[t]
\vskip 4.8truecm
\includegraphics{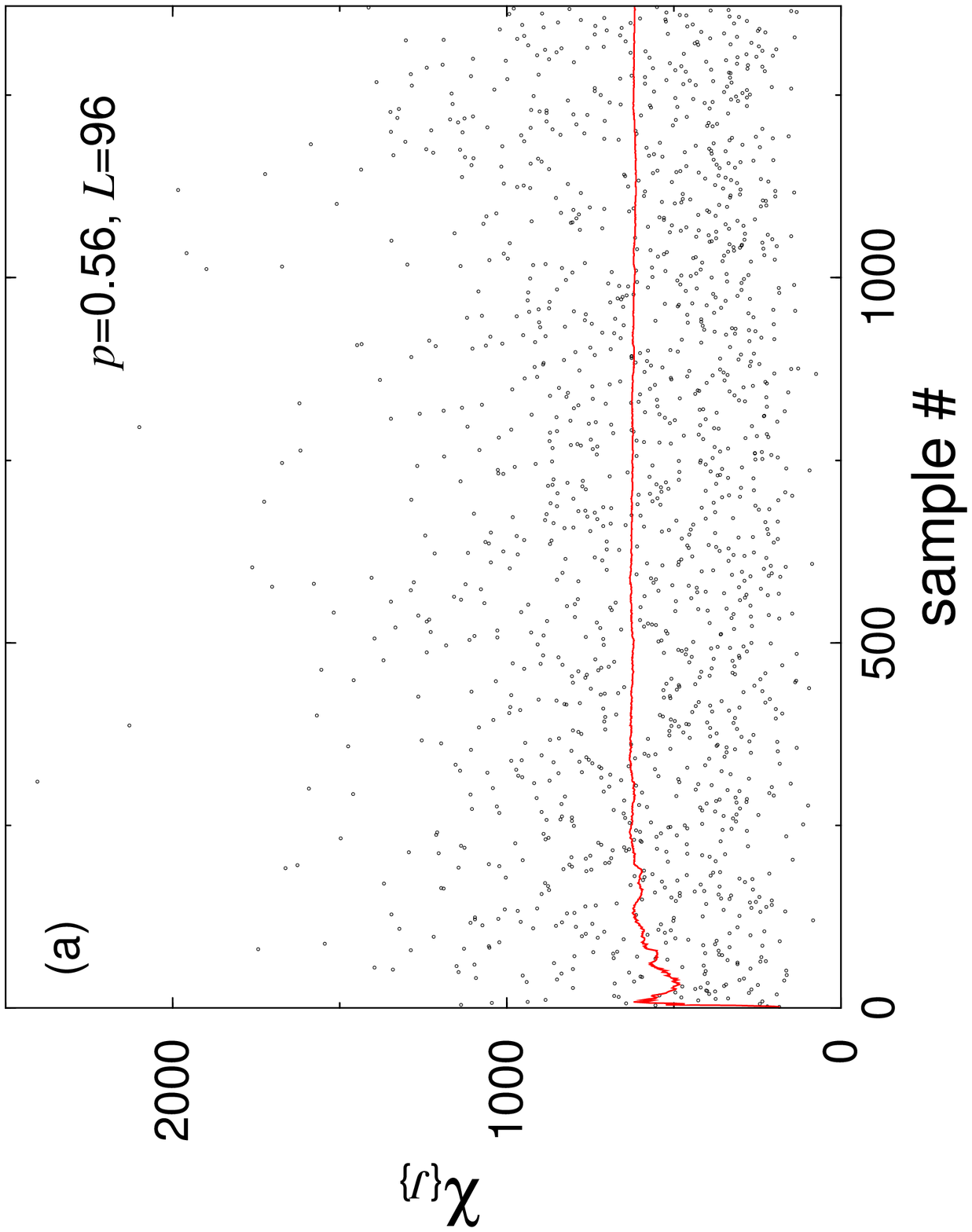}
\includegraphics{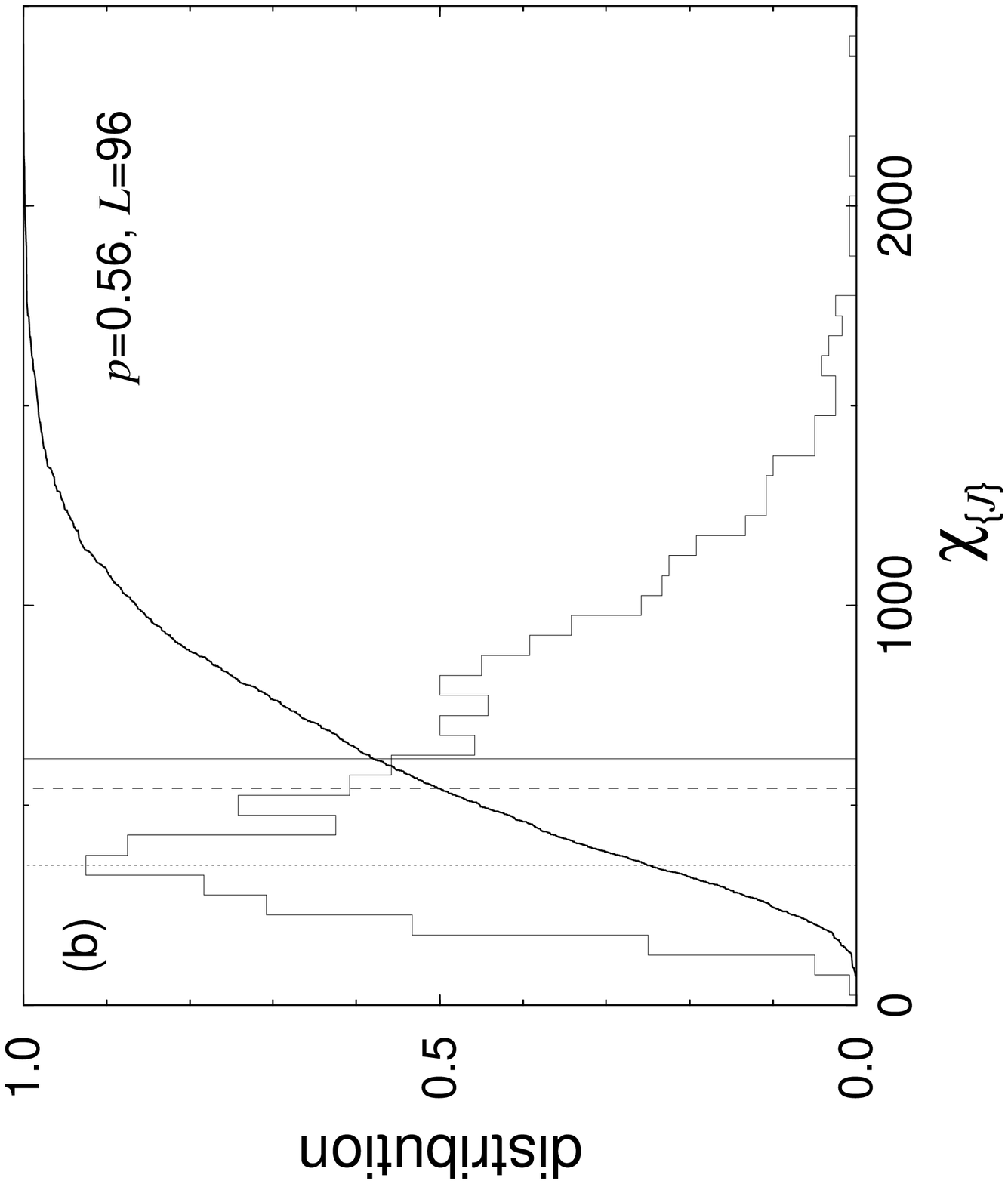}
\caption{\label{fig:chi_p056_L96} Distribution of the magnetic susceptibility
$\chi_{\{J\}}$ for $p=0.56$ and $L=96$, as obtained at a temperature close to 
the point where the average $[\chi_{\{J\}}(T)]_{\rm av}$ has its maximum.
The solid line in (a) is the running average, and the vertical lines in (b) 
show from left to right the most probable, median, and mean value.}
\end{figure}

Recent Monte Carlo (MC) computer simulations of the 3D {\em site\/}-diluted 
3-state \cite{Ballesteros00} and {\em bond\/}-diluted 4-state \cite{Chatelain01}
Potts models, which both exhibit first-order phase transitions in the pure 
case, have confirmed this
expectation numerically. For a complementary study using high-temperature 
series expansions, see Ref.~\cite{meik2}. One important issue in MC studies
is the number of independent disorder realizations that are required for 
reliable quenched averages. An answer to this question can usually only be
given a posteriori, after having already estimated the associated distributions
of thermodynamic quantities such as the
susceptibility, since these distributions can be highly asymmetric.
As is illustrated in Fig.~\ref{fig:chi_p056_L96}, one result of our large-scale 
simulations \cite{Chatelain01} is that this asymmetry is
particularly pronounced in the softening regime. In what follows we shall
further illustrate and elucidate this effect and its consequences with a few 
selected examples. 

\section{Model and Simulation Setup}
\label{sec:2}

The 3D bond-diluted $q$-state Potts model is defined by the Hamiltonian
    \begin{equation}
      H= - \sum_{\langle ij \rangle} J_{ij}\delta_{\sigma_i,\sigma_j};
      \quad \sigma_i=1, \ldots, q,
      \label{eq1}
    \end{equation}
where the sum extends over all pairs of neighbouring sites on a cubic lattice
of size $L^3$ with periodic boundary conditions, and the couplings $J_{ij}$ 
are chosen according to the distribution
    \begin{equation}
      \wp(J_{ij})=p\ \!\delta(J_{ij}-J)+(1-p)\ \!\delta(J_{ij}),
      \label{eq2}
    \end{equation}
where $p$ is the concentration of magnetic bonds such that $p=1$ 
corresponds to the pure case where the 4-state model exhibits a rather strong
first-order transition at $K_c(1) \equiv K_c = J/(k_B T_c) = 0.62863(2)$.

We simulated the diluted  model in the regime $p_c < p < p_t$ of second-order 
transitions, where $p_c = 0.248\,812\,6(5)$ is the bond-percolation 
threshold \cite{Lorenz1998} and $p_t \approx 0.80 $ denotes the tricritical 
concentration \cite{Chatelain01},
with the Swendsen-Wang cluster algorithm \cite{SW87}, and 
in the regime $p_t < p \le 1$ of first-order transitions with the
multibondic method \cite{mubo}. The numerically determined phase diagram 
in the dilution-temperature plane was found to agree very well with the
single-bond effective-medium (EM) approximation \cite{turban},
\begin{equation}
K^{\rm EM}_c (p)=\ln \left[(1-p_c ) e^{K_c(1)}-(1-p)\over p-p_c\right],   
\label{eq:EM}
\end{equation}
where $K_c(1)$ is the transition point 
of the pure system and $p_c$ the percolation threshold given 
above. Our analysis of both the autocorrelation time and the interface 
tension \cite{Chatelain01}
led to the conclusion of a tricritical point around $p=0.80$.

To arrive at these results, for each dilution, temperature and lattice
size, the MC estimates $\langle Q_{\{J\}} \rangle$ of thermodynamic quantities 
$Q_{\{J\}}$ for a given random distribution $\{J\}$ of diluted bonds were 
averaged over 2\,000 -- 5\,000 disorder realizations,
\begin{equation}
Q = [\langle Q_{\{J\}} \rangle]_{\rm av} = \int\! {\cal D}J_{ij}\wp(J_{ij})
\langle Q_{\{J\}} \rangle  = \int\! d \langle Q_{\{J\}} \rangle 
{\cal P}(\langle Q_{\{J\}}\rangle)
\langle Q_{\{J\}} \rangle, 
\end{equation}
where ${\cal P}(\langle Q_{\{J\}}\rangle)$ denotes the empirically determined 
distribution of $\langle Q_{\{J\}}\rangle$ discussed in the next section.

\begin{figure}[t]
\begin{center}
\includegraphics[angle=-90,scale=0.32]{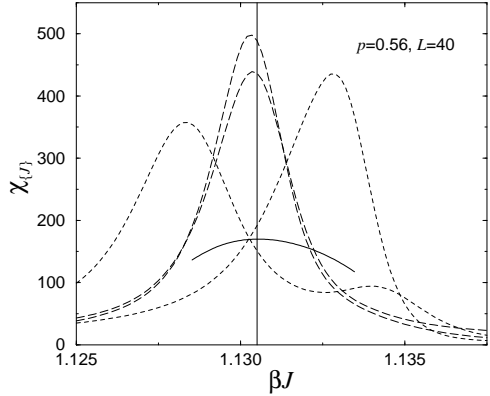}
\end{center}\vspace*{-0.4cm}
\caption{\label{fig:rare_chi}
Temperature variation of four different realizations at $p=0.56$ and $L=40$
(long and short dashed lines).
The solid line with a maximum around $\chi_{\{J\}} \approx 160$
shows the average susceptibility $\chi = [\chi_{\{J\}}]_{\rm av}$.
}
\end{figure}

\section{Distribution over Disorder Realizations}
\label{sec:3}

In Fig.~\ref{fig:chi_p056_L96} the distribution of susceptibility measurements
at a temperature close to the peak of its disorder average is shown 
for $p=0.56$ and $L=96$, i.e., for a dilution in the centre of the softening
regime $0.25 \approx p_c < p < p_t \approx 0.80$. The solid line 
Fig.~\ref{fig:chi_p056_L96}(a) 
is the running average which, despite the quite asymmetric distribution,
appears rather stable from about 400 realizations on. The resulting
probability density (histogram) and distribution (accumulated density) are
depicted in Fig.~\ref{fig:chi_p056_L96}(b). By the vertical solid and 
dashed line we have also indicated the average and median value, respectively. 
The dotted vertical line at about $\chi_{\{J\}} = 350$, which perfectly 
coincides with the most probable value, has been computed by averaging over 
only those $50\%$ of the realizations whose $\chi_{\{J\}}$-value is smaller 
than the median value.

In Fig.~\ref{fig:rare_chi} we illustrate how the average susceptibility
emerges from the individual contributions of each disorder realization.
The location of its
maximum is indicated by the vertical line. Two events corresponding
to large values of $\chi_{\{J\}}$ (rare events) are shown as long dashed
lines, and two events with small values of $\chi_{\{J\}}$ (typical events)
are shown as short dashed lines.

\begin{figure}[t]
\begin{center}
\includegraphics[scale=0.35]{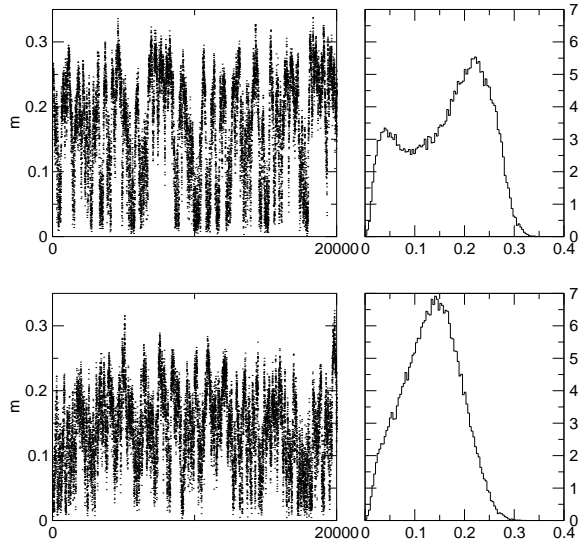}
\end{center}
\caption{\label{fig:thermalize}
Time evolution of the magnetization for measurements close to the (average)
susceptibility peak for two different realizations at $p=0.56$ and $L=40$,
and the associated thermal equilibrium distributions.
}
\end{figure}

Since each $\chi_{\{J\}}$ reflects the thermal fluctuations in the 
magnetization $m$,
we also had a closer look at the time evolution of $m$ and the associated
equilibrium distribution. Two examples are shown in Fig.~\ref{fig:thermalize}, 
where the upper part corresponds to a rare event with large
susceptibility and the lower one to a typical event with smaller
magnetization fluctuations. The first realization still behaves 
first-order like as can be seen more clearly 
on the right part of the figure, where the resulting thermal
probability density of the magnetization exhibits a double
peak.

\section{Conclusions}
\label{sec:4}

By performing large-scale Monte Carlo simulations
we have investigated the influence of bond dilution on the thermodynamic
properties of the 3D 4-state Potts model. For strong dilution with
$p < 0.80$ we obtained clear evidence for softening to a continuous 
transition. Here we have focused on a rather detailed study of the 
thermalization properties of individual realizations and the disorder 
distribution of thermodynamic quantities such as the susceptibility.
Among other observations this reveals that in the softening regime 
a few rare realizations may still exhibit signals reminiscent of a first-order
transition, e.g., a double peak in the magnetization histogram. On the
one hand, this gives rise to rare events with large susceptibilities leading
to quite asymmetric distributions over the disorder, and on the other
hand this complicates the technical issue of deciding which Monte Carlo
update algorithm is best suited for the problem at hand.  

\section*{Acknowledgements}
\label{sec:5}

We are grateful to Meik Hellmund and Loic Turban for helpful discussions.
Work partially supported by a PROCOPE collaborative grant of DAAD and EGIDE,
the EU network ``EUROGRID: {\em Discrete Random Geometries: From Solid State 
Physics to Quantum Gravity\/}", the German-Israel-Foundation (GIF)
grant No.~I-653-181.14/1999, and the computer-time grants 2000007 of 
CRIHAN, hlz061 of NIC J\"ulich, and h0611 of LRZ M\"unchen.

%

\begin{thebibliography}{99.}
%
%
%
%
%
%

\bibitem{Folk2001} 
R. Folk, Y. Holovatch, T. Yavors'kii:
Physics Uspiekhi {\bf 173}, 175 (2003) [e-print cond-mat/0106468]

\bibitem{Harris74} 
A.B. Harris: J. Phys. C \textbf{7}, 1671 (1974)
        
\bibitem{ImryWortis79} 
Y. Imry, M. Wortis: Phys. Rev. B \textbf{19}, 3580 (1979)
 
\bibitem{Aizenman89} 
M. Aizenman, J. Wehr: Phys. Rev. Lett. \textbf{62}, 2503 (1989)

\bibitem{BeCh02}
B. Berche, C. Chatelain: e-print cond-mat/0207421

\bibitem{cardy} 
J. Cardy, J.L. Jacobsen: Phys. Rev. Lett. \textbf{79}, 4063 (1997)

\bibitem{cardy_statphys}
For a review, see J. Cardy: Physica A \textbf{263}, 215 (1999)

\bibitem{Ballesteros00} 
H.G. Ballesteros, L.A. Fern\'andez, V. Mart\'\i n-Mayor, 
A. Mu\~noz Sudupe, G. Parisi, J.J. Ruiz-Lorenzo:
Phys. Rev. B \textbf{61}, 3215 (2000)

\bibitem{Chatelain01}
C. Chatelain, B. Berche, W. Janke, P.-E. Berche:
Phys. Rev. E \textbf{64}, 036120 (2001);
%
C. Chatelain, P.-E. Berche, B. Berche, W. Janke:
Nucl. Phys. B (Proc. Suppl.) \textbf{106\&107}, 899 (2002);
%
Comp. Phys. Comm. \textbf{147}, 431 (2002) 

\bibitem{meik2} 
M. Hellmund, W. Janke:
Nucl. Phys. B (Proc. Suppl.) \textbf{106\&107}, 923 (2002);
Phys. Rev. E \textbf{67}, 026118 (2003)

\bibitem{Lorenz1998} 
C.D. Lorenz, R.M. Ziff: Phys. Rev. E \textbf{57}, 230 (1998)

\bibitem{SW87} 
R.H. Swendsen, J.S. Wang: Phys. Rev. Lett. \textbf{58}, 86 (1987)

\bibitem{mubo}
W. Janke, S. Kappler: Phys. Rev. Lett. \textbf{74}, 212 (1995)

\bibitem{turban} 
L. Turban: Phys. Lett. A \textbf{75}, 307 (1980); 
J. Phys. C \textbf{13}, L13 (1980)

\end{thebibliography}
%


\clearpage
\addcontentsline{toc}{section}{Index}
\flushbottom
\printindex

\end{document}